%% file: 67329_2.tex
\documentclass[fp,twocolumn]{jpsj3}
\usepackage{bm}
\usepackage{comment}
\usepackage[dvipdfmx]{graphicx}
\usepackage{graphicx}
\usepackage{color}
\usepackage{overpic}
\newcommand{\barT}[1]{{\cal T}_{\overline{#1}}}
\newcommand{\barR}[1]{{\cal R}_{\overline{#1}}}
\newcommand{\Ttilde}[1]{\tilde{{\cal T}}_{#1}}
\newcommand{\Rtilde}[1]{\tilde{{\cal R}}_{#1}}
\newcommand{\barTtilde}[1]{\tilde{{\cal T}}_{\overline{#1}}}
\newcommand{\barRtilde}[1]{\tilde{{\cal R}}_{\overline{#1}}}

\title{
Dephasing-Induced Stabilization
of a Perfectly Conducting Channel in Disordered Graphene Nanoribbons
with Zigzag Edges
}

\author{
Yuji Shimomura and Yositake Takane
}

\inst{
Department of Quantum Matter, Graduate School of Advanced Sciences of Matter,\\
Hiroshima University, Higashihiroshima, Hiroshima 739-8530, Japan
}

\abst{
Electron transport in a disordered graphene nanoribbon with zigzag edges is
crucially affected by a perfectly conducting channel (PCC),
which  is stabilized if intervalley scattering is ignorable.
In the presence of such a PCC, the dimensionless conductance $g$
of the system decreases to the quantized value of $g = 1$
with increasing system length $L$.
In the realistic case where intervalley scattering is weak but not ignorable,
the PCC is gradually destabilized with increasing $L$,
and $g$ eventually decays to zero owing to the onset of Anderson localization.
Here, we show that
such destabilization of the PCC can be relaxed by pure dephasing.
We numerically calculate $g$ in the presence of
long-range impurities, which induce weak intervalley scattering,
taking the dephasing effect into account.
It is demonstrated that, under sufficient dephasing,
the decay of $g$ in the regime of $g \lesssim 1$ is strongly suppressed
and the quasi-quantization of $g$ (i.e., $g \sim 1$)
can be observed in a wide region of $L$.
}

\begin{document}
\maketitle
\section{Introduction}

A perfectly conducting channel (PCC) designates a conducting channel that
perfectly transmits an electron from one end to the other
in spite of the presence of disorder.
The most well-known example is a one-dimensional (1D) chiral edge channel
of two-dimensional (2D) quantum Hall insulators.~\cite{buttiker}
This 1D edge channel needs a strong magnetic field for its stabilization.
So far, PCCs have been shown to appear in various carbon nanostructures
and topological insulators under no external field.

Ando and co-workers~\cite{ando1,ando2,nakanishi,ando3,ando4} showed
that a PCC appears in disordered carbon nanotubes (CNTs) with a gapless spectrum
in the absence of intervalley scattering.
The requirement of no intervalley scattering is explained as follows.
A CNT possesses two energy valleys in the reciprocal space,
and the effective Hamiltonian describing each valley is invariant
under a time-reversal operation $\cal{T}$ that satisfies ${\cal T}^2 = -1$;
the subsystems corresponding to each valley have the symplectic symmetry.
Furthermore, the number of conducting channels in each valley is always odd
regardless of the Fermi level.
The existence of a PCC is guaranteed by the symplectic symmetry combined with
an odd number of conducting channels.
If intervalley scattering occurs
and the two valleys are coupled as its consequence,
these two conditions break down and hence the PCC disappears.
In CNTs, an impurity potential with a range larger
than the lattice constant induces only very weak intervalley scattering
since the two energy valleys are well separated in the reciprocal space.
Thus, we expect that the above two conditions are approximately satisfied
if a CNT contains only such long-range impurities (LRIs),
leading to the appearance of a PCC.
Clearly, there is no PCC in the presence of short-range impurities (SRIs).

A 1D helical edge channel of
2D quantum spin-Hall insulators~\cite{kane1,kane2,onoda,bernevig,qi,murakami}
can be regarded as a typical example of a PCC in topological insulators.
Its protection mechanism against disorder is essentially equivalent to
that of a PCC in CNTs.
However, as quantum spin-Hall insulators typically possess only a single valley,
the disturbance due to intervalley scattering is irrelevant in this case.
A similar PCC is stabilized in three-dimensional (3D) weak topological
insulators in various situations.~\cite{ran,imura,ringel,yoshimura,kobayashi}
In 3D strong topological insulators,
a PCC can appear only when
a $\pi$ magnetic flux penetrates the bulk of a sample
without touching surface states.~\cite{zhang,egger,bardarson}

Wakabayashi and co-workers~\cite{wakabayashi1,wakabayashi2,wakabayashi3}
showed that disordered graphene nanoribbons with zigzag edges accommodate
a PCC, on which our interest is focused in this paper.
Hereafter, a graphene nanoribbon with zigzag edges is simply
referred to as a zigzag nanoribbon.
As in the case of CNTs, zigzag nanoribbons possess
two energy valleys in the reciprocal space.
The important feature of zigzag nanoribbons is that conducting channels
are imbalanced between the two propagating directions in each valley.
That is, the number of conducting channels going in one direction
is one greater or smaller than
that going in the other direction,
regardless of the Fermi level.
This directly results in the stabilization of
a PCC~\cite{barnes1,barnes2,hirose} if intervalley scattering is ignorable.
Thus, we expect the appearance of a PCC in zigzag nanoribbons
containing only LRIs.
In contrast to the case of CNTs,
the symmetry of the system plays no role in this case.
It has been shown that disordered graphene nanoribbons with armchair edges
also accommodate a similar PCC.~\cite{yamamoto,wurm}

If a PCC stably exists, the dimensionless conductance $g$ of the system
decreases to the quantized value of $g = 1$
with increasing system length $L$.
In CNTs and graphene nanoribbons, we expect the appearance of
a PCC only when the disorder of the system is long-range, as noted above.
Indeed, if the spatial range of disorder is sufficiently large,
the strength of intervalley scattering becomes very weak.
However, in actual situations,
it is impossible to completely suppress intervalley scattering.
The effect of residual intervalley scattering gradually manifests itself
with increasing $L$ and eventually destabilizes a PCC.
In this case, the behavior of $g$ may not be distinguishable from that in
an ordinary system with no PCC.

The effect of inelastic scattering may be another important obstacle to the observation of a PCC.
Inelastic scattering caused
by electron--electron and/or electron--phonon interactions
affects low-energy electrons mainly through energy relaxation and dephasing.
At low temperatures,
pure dephasing most significantly influences the transport properties.
Indeed, dephasing directly destabilizes a PCC in CNTs as well as topological insulators since it weakens the underlying symplectic symmetry of the system,
except for the case with only a single channel, where it becomes a PCC and is relatively robust against dephasing.~\cite{ando3,ando4}
Contrastingly, a PCC in zigzag nanoribbons relies on no symmetry of the system, so dephasing does not necessarily disturb it.
Previous studies~\cite{takane2,takane3} based on the Boltzmann transport equation
indicate that a PCC remains even in the incoherent limit.
However, it is not clear how the behavior of a PCC changes
with the reduction of phase coherence.

In this paper, the effect of pure dephasing on a PCC in zigzag nanoribbons
is studied by numerical simulations of the dimensionless conductance.
We show that dephasing does not disturb a PCC
but rather relaxes its destabilization due to weak intervalley scattering.
To clarify the dephasing effect on the PCC
we numerically calculate the average dimensionless conductance
$\langle g \rangle$ in zigzag nanoribbons with LRIs at zero temperature,
taking account of dephasing within the model presented in Ref.~\citen{ando3}.
In the case without dephasing, we observe that $\langle g \rangle$ rapidly decreases
to the quantized value of $\langle g \rangle = 1$ with increasing $L$,
implying the presence of a PCC,
and then it exponentially decays below the quantized value,
reflecting the destabilization of the PCC.
%and then it exponentially decays below the quantized value toward zero
%owing to the onset of Anderson localization.
However, in the presence of sufficiently strong dephasing, we observe that
the exponential decay of $\langle g \rangle$ in the regime of
$\langle g \rangle \lesssim 1$ is significantly relaxed owing to the suppression of Anderson localization.
%$\langle g \rangle < 1$ is strongly suppressed
%and the quasi-quantization of $\langle g \rangle$
Consequently, the quasi-quantization of $\langle g \rangle$
(i.e., $\langle g \rangle \sim 1$) can be observed in a wide region of $L$.
This result should encourage experimental attempts to detect
a PCC in zigzag nanoribbons.~\cite{baringhaus}

In the next section, we present the tight-binding model
for a zigzag nanoribbon.
We assume that every impurity potential distributed over the system
is described by a Gaussian form of spatial range $d$.
This corresponds to an LRI (SRI)
when $d$ is larger (smaller) than the lattice constant $a$.
We compute the dimensionless conductance using the Landauer formula
by numerically determining the scattering matrix for the system.
The model for describing the pure dephasing is also introduced.
In Sect.~3, the numerical results of the average dimensionless conductance
are presented for the case with LRIs and that with SRIs.
We observe that the destabilization of the PCC
can be relaxed by the dephasing in the former case.
In Sect.~4, the numerical results of the previous section are compared with
an analytical expression for the dimensionless conductance
derived from the Boltzmann transport equation.~\cite{takane2}
We see that the numerical result in the case with strong dephasing
is accurately fitted by the analytical result,
implying that our model appropriately describes the effect of dephasing.
The last section is devoted to summary and conclusion.
Preliminary results of this work have been briefly reported
in Ref.~\citen{ashitani}.

\section{Model and Formulation}

We consider a zigzag nanoribbon consisting of $M$ zigzag lines
placed along the $x$-axis (see Fig.~1).
Its band structure is shown in Fig. \ref{fig:zzband} in the case of $M = 30$.
One can see that in the left (right) valley, the number of right-going
(left-going) channels is one greater than that of left-going
(right-going) channels regardless of the location of the Fermi level.
This indicates that a right-going PCC appears in the left valley while
a left-going PCC appears in the right valley.~\cite{wakabayashi1}

\begin{figure}[htpb]
	\centering
		\includegraphics{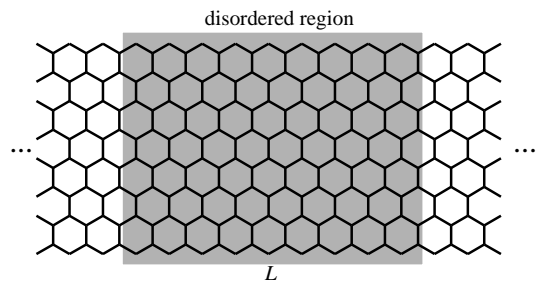}
	\caption{
	Illustration of zigzag nanoribbon.
%	$a$ is a lattice constant
%	and $N$ is a ribbon width.
%	determines the transport properties.
	The gray area of length $L$
	represents the disordered region with
	randomly distributed impurities. 
%	have the Gaussian form as the disordered zone.
	The left and right regions without disorder
	are regarded as perfect leads.	
	}
	\label{fig:zigzag ribbon}
\end{figure}
\begin{figure}
	\begin{overpic}
		{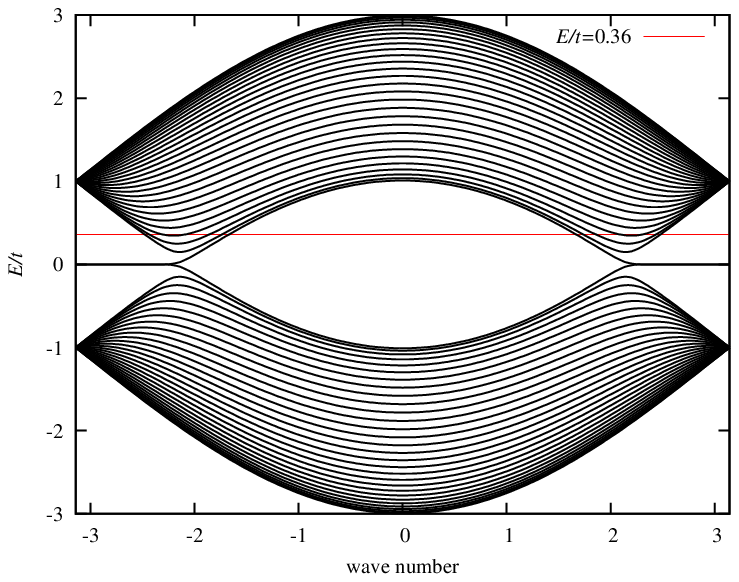}
		\put(56,11){\includegraphics[scale=0.1]{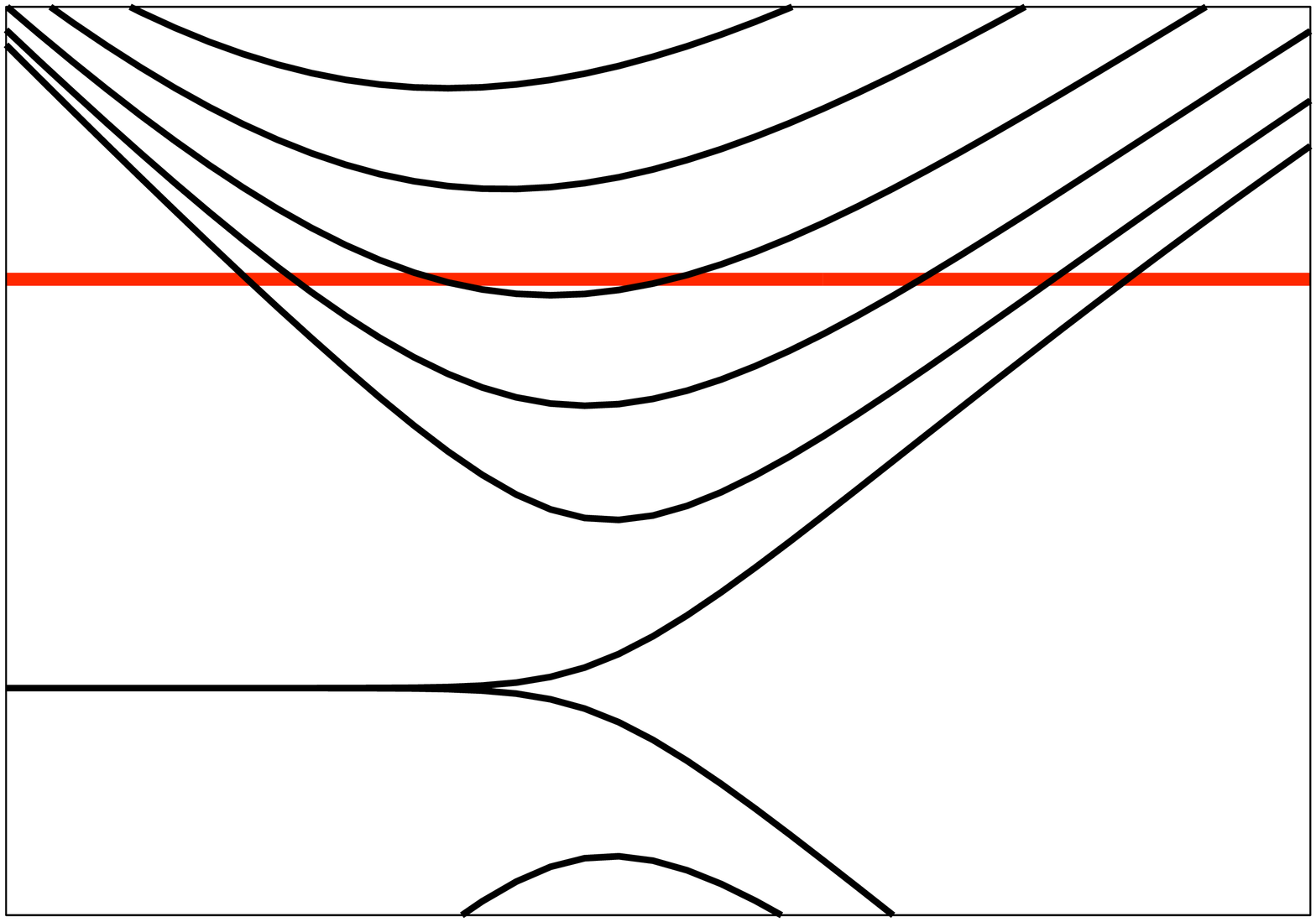}}
	\end{overpic}
	\caption{
	Band structure of zigzag nanoribbon with $M=30$.
	The inset represents the magnification of subbands
	in the left valley.
	}
	\label{fig:zzband}
\end{figure}

\subsection{Hamiltonian}
We describe $\pi$ electrons in zigzag nanoribbons
by using the nearest-neighbor tight-binding model
\begin{align}
	H=-t \sum_{n. n.} |i\rangle \langle j|
	+ \sum_{i} V({\bm r}_{i})|i\rangle \langle i|,
\end{align}
where $t$ is the hopping integral between neighboring sites,
$|i\rangle$ is the $\pi$ orbital on site $i$,
and $V({\bm r}_{i})$ is the impurity potential
with ${\bm r}_i$ being the position of site $i$.
We randomly distribute impurities
in a region of length $L$ (see Fig. \ref{fig:zigzag ribbon}).
We assume that
each site is occupied by an impurity with probability $P$
and the potential of each impurity is characterized by
a Gaussian form of spatial range $d$.
Hence, $V({\bm r}_i)$ is represented as
\begin{align}
	V({\bm r}_i)=\sum_{j} u({\bm r}_{j})\text{exp}\left(-\frac{|{\bm r}_i-{\bm r}_j|^2}{d^2}\right),
	\label{}
\end{align}
where $u({\bm r}_j)$ is the strength of an impurity at site $j$.
We assume that $u({\bm r}_j)$ is uniformly distributed within $|u|<u_{\rm{max}}/2$.
Note that the degree of disorder is determined
by $d$, $P$, and $u_{\rm{max}}$.

\subsection{Scattering matrix}
The electron transport property of a zigzag nanoribbon is determined
by the scattering matrix,
which consists of transmission matrices $\bm{t}$ and $\bm{t}'$
and reflection matrices $\bm{r}$ and $\bm{r}'$,
where $\bm{t}$ and $\bm{r}$ ($\bm{t}'$ and $\bm{r}'$)
describe the scattering of an electron incoming from the left (right).
Let $N_{\rm c}$ be the number of conducting channels 
for a given value of the Fermi energy $E$.
The dimensions of the transmission and reflection matrices are $N_{\rm c} \times N_{\rm c}$.
In calculating the scattering matrix,
we consider that the left and right of the disordered region serve as
perfect leads of semi-infinite length without disorder.
The transmission and reflection matrices
can be obtained by using a recursive Green's function method.
The dimensionless conductance at zero temperature is obtained 
using the Landauer formula $g(E)=\text{tr}({\bm t}^{\dagger}{\bm t})$,
where ${\bm t}^{\dagger}$ is the Hermitian conjugate of $\bm{t}$.

\subsection{Dephasing}

Generally speaking, dephasing suppresses quantum effects,
particularly quantum interference effects,
and tends to reveal classical behaviors of electrons in some cases.

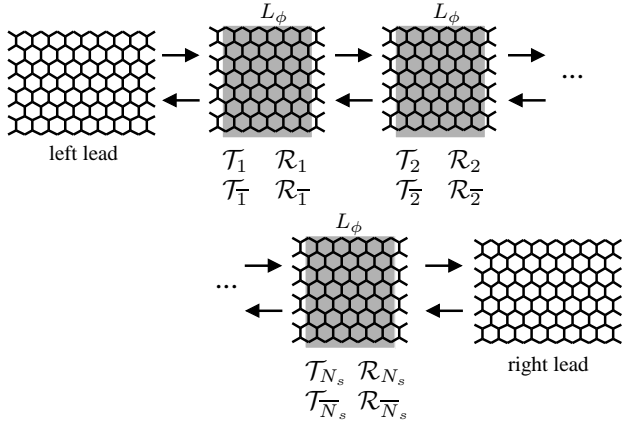
\begin{figure}[htpb]
	\begin{center}
		\input{67329Fig3.pstex_t}
	\end{center}
	\caption{
	Hypothetical decomposition of the disordered region
	of length $L$ into $N_s$ segments of equal length $L_\phi$.
	We assume that the phase coherence of electrons is lost across
	adjacent segments.
	}
	\label{fig:dephasing}
\end{figure}

In order to incorporate this effect into our model,
we hypothetically decompose the disordered region into $N_s$ segments 
of equal length $L_\phi$, as shown in Fig.~\ref{fig:dephasing},
and assume that the phase coherence of electrons is 
lost across adjacent segments while in each segment
the phase coherence is completely preserved.\cite{ando3}
Hence, $L_\phi$ can be regarded as the phase coherence length.
%Once the scattering matrix of each segment is given,
%we can obtain the total transmission probability of the whole system by
%requiring
We require the continuity of the charge current,
instead of the continuity of a wave function,
in each channel between
adjacent segments.
With this procedure, the phase coherence of electrons completely breaks
across adjacent segments.
Let us express the transmission and reflection matrices
for the $n$th segment as
${\bm t}_n$, ${\bm r}_n$, ${\bm t}'_n$, and ${\bm r}'_n$.
In terms of them,
the transmission probability matrix ${\cal T}_n$ ($\barT{n}$)
and the reflection probability matrix ${\cal R}_n$ ($\barR{n}$)
for an electron incoming from the left (right)
are defined as
\begin{align}
	\begin{aligned}
	\left[{\cal T}_n\right]_{\alpha\beta}&=|[{\bm t}_n]_{\alpha\beta}|^2,\\
	[{\cal R}_n]_{\alpha\beta}&=|[{\bm r}_n]_{\alpha\beta}|^2,\\
	[\barT{n}]_{\alpha\beta}&=|[{\bm t}_n']_{\alpha\beta}|^2,\\
	[\barR{n}]_{\alpha\beta}&=|[{\bm r}_n']_{\alpha\beta}|^2.
	\end{aligned}
	\label{aaaa}
\end{align}
For the system constituted
by combining the first, second, $\cdots$, $n$th segments in series,
we define $\Ttilde{n}$ ($\barTtilde{n}$) and $\Rtilde{n}$ ($\barRtilde{n}$)
as the transmission and reflection probability matrices
for an electron incoming from the left (right), respectively.
The continuity of the charge current ensures that
they obey
the following recursive relation.\cite{ando3}
\begin{align}
	&\begin{pmatrix}
		\Rtilde{n+1} & \barTtilde{n+1}\\
		\Ttilde{n+1} & \barRtilde{n+1}
	\end{pmatrix}
	=
	\begin{pmatrix}
		\Rtilde{n} & 0\\
		0 & \barR{n+1}
	\end{pmatrix} \nonumber\\
	&+
	\begin{pmatrix}
		0 & \barTtilde{n}\\
		{\cal T}_{n+1} & 0
	\end{pmatrix}
	\begin{pmatrix}
		1 & -\barRtilde{n}\\
		-{\cal R}_{n+1} & 1
	\end{pmatrix}^{-1}
	\begin{pmatrix}
		\Ttilde{n} & 0\\
		0 & \barT{n+1}
	\end{pmatrix},
%	\label{mat4}
\end{align}
with $\Ttilde{1}={\cal T}_{1}$, $\Rtilde{1}={\cal R}_{1}$,
$\barTtilde{1}=\barT{1}$, and $\barRtilde{1}=\barR{1}$.
The dimensionless conductance is given by
\begin{align}
	g=\sum^{N_{\rm c}}_{\alpha, \beta=1}\left[\Ttilde{N_s}\right]_{\beta\alpha}
	\label{}
\end{align}
in the presence of dephasing.

\section{Numerical Results}
We separately consider 
the case with LRIs and that with SRIs.
In the former case, we expect that a PCC appears
as intervalley scattering is sufficiently weak,
while strong intervalley scattering forbids its appearance in the latter case.
These two cases are realized by appropriately choosing
$d$.
With $P=0.1$, we set $d/a=1.5$ and $u_\text{max}/t=0.1$ in the LRI case,
and $d/a=0.05$ and $u_\text{max}/t=1.0$ in the SRI case.
We fix $M=30$ and $E/t=0.36$, resulting in $N_{\text c}=7$, in the following calculations.

\begin{figure}[htpb]
	\centering
	\includegraphics{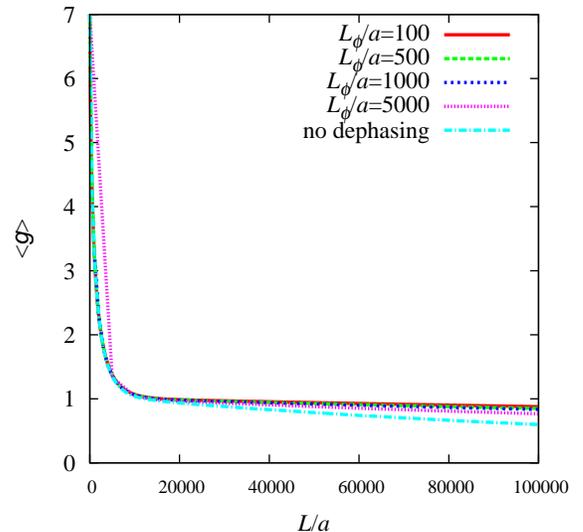}
	\caption{
	(Color online)
	Average dimensionless conductance $\langle g\rangle$
	in the LRI case
	for several values of $L_\phi/a$.
	}
	\label{fig:LRI}
\end{figure}

\begin{figure}[htpb]
	\centering
	\includegraphics{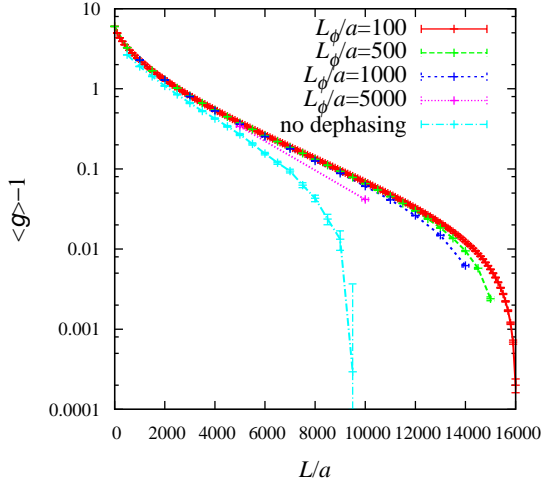}
	\caption{
	(Color online)
	Semilog plot of $\langle g\rangle -1$ in the case of LRI.
	In a certain region of $L/a$,
	$\langle g\rangle$ decreases exponentially 
	toward 1 as a function of $L/a$.
	}
	\label{fig:g_minus}
\end{figure}

\begin{figure}[htpb]
	\centering
	\includegraphics{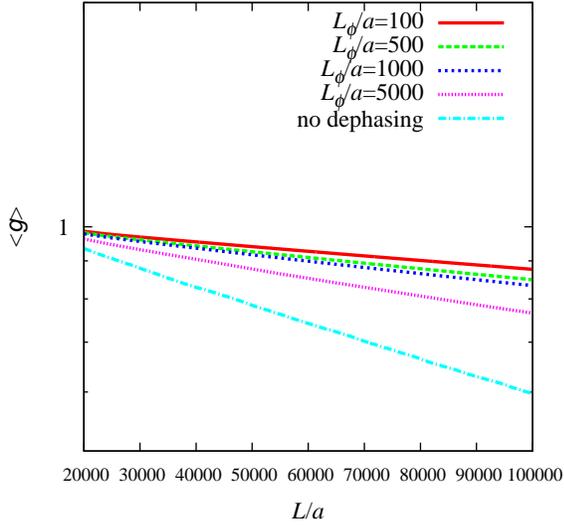}
	\caption{
	(Color online)
	Semilog plot of $\langle g\rangle$ in the LRI case
	in the regime of $\langle g\rangle<1$.
	$\langle g\rangle$ decays exponentially as a function
	of $L/a$ even in the presence of dephasing.
	}
	\label{fig:LRI_log}
\end{figure}

Let us consider the LRI case. 
We use 2000 samples with different impurity configurations 
to calculate the average dimensionless conductance $\langle g\rangle$.
Figure \ref{fig:LRI} shows
that $\langle g\rangle$ rapidly decreases to 1 with increasing $L/a$
and then the decrease becomes very slow once it decays below 1,
indicating the presence of a PCC.
In the regime of $\langle g\rangle\lesssim 1$,
the decay of $\langle g\rangle$ becomes slower
as $L_\phi$ becomes smaller.
This can be explained by considering that
the decay of $\langle g\rangle$ below 1 is
accelerated by Anderson localization.
The effect of Anderson localization
is suppressed by dephasing and
hence the decay of $\langle g\rangle$ is also suppressed
with decreasing $L_\phi$.
This indicates that dephasing indirectly stabilizes a PCC,
although dephasing itself does not weaken intervalley scattering.
We observe the quasi-quantization of $\langle g\rangle$
(i.e., $\langle g\rangle\sim1$) 
in a wide region of $L/a$ when $L_\phi$ is sufficiently small.
Figure~\ref{fig:g_minus} shows a semilog plot of $\langle g\rangle -1$.
Error bars at each data point represent $(\text{var}\{g\}/N_\text{sam})^{1/2}$,
where
$\text{var}\{g\}=\langle g^2\rangle-\langle g\rangle^2$ and
$N_\text{sam}$ is the number of samples used to calculate the average.
We find that $\langle g\rangle$ decreases exponentially toward 1
in a certain region of $L/a$ in all cases.
In the case without dephasing,
this is in accordance with existing random matrix theory.\cite{takane1}
Figure~\ref{fig:LRI_log} shows a semilog plot of $\langle g\rangle$
in the regime of $\langle g\rangle<1$.
We find that $\langle g\rangle$ decays exponentially
regardless of $L_\phi$.
In the absence of dephasing,
it is natural that $\langle g\rangle$ decays exponentially,
reflecting the onset of Anderson localization.
The exponential decay of $\langle g\rangle$ even
in the presence of dephasing
should be regarded as a characteristic feature of the system
in which conducting channels are imbalanced
between two propagating directions.\cite{takane2}

\begin{figure}[htpb]
	\centering
		\includegraphics{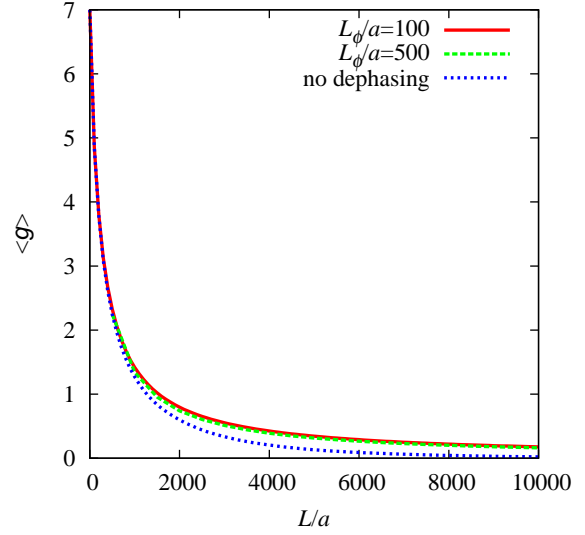}
	\caption{
	(Color online)
	Average dimensionless conductance $\langle g\rangle$
	in the SRI case.
	}
	\label{fig:SRI}
\end{figure}

\begin{figure}[htpb]
	\centering
		\includegraphics{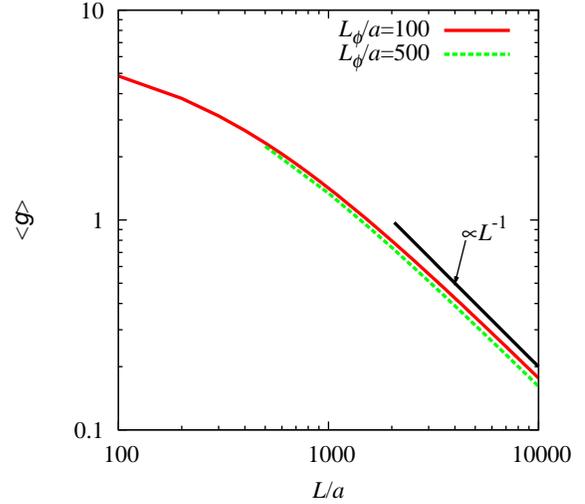}
	\caption{
	(Color online)
	Log-log plot of $\langle g\rangle$ in the
	SRI case with dephasing.
	$\langle g\rangle$ asymptotically becomes
	inversely proportional to $L/a$.
	}
	\label{fig:SRI_log-log}
\end{figure}

\begin{figure}[htpb]
	\centering
		\includegraphics{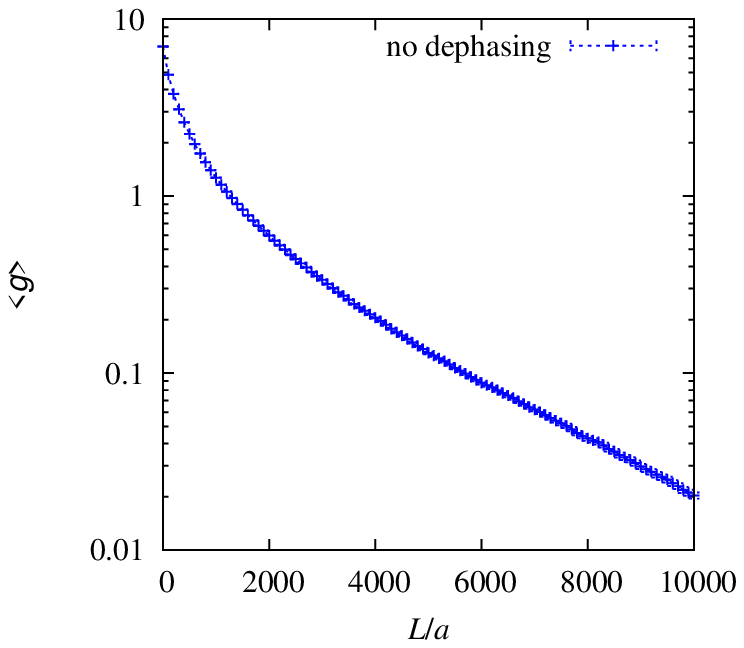}
	\caption{
	(Color online)
	Semilog plot of $\langle g\rangle$ in the
	SRI case with no dephasing.
	$\langle g\rangle$ decays exponentially
	as a function of $L/a$.
	}
	\label{fig:SRI_semilog}
\end{figure}

We turn to the SRI case.
We use 10000 samples with different impurity configurations 
to calculate the average dimensionless conductance.
Figure \ref{fig:SRI} shows 
that
$\langle g\rangle$ decays to zero with increasing $L/a$,
indicating the absence of a PCC.
Figure \ref{fig:SRI_log-log} shows a log-log plot of $\langle g\rangle$
in the presence of dephasing.
We find that $\langle g\rangle$ asymptotically becomes inversely proportional to $L/a$,
manifesting that Ohm's law is satisfied upon
the suppression of Anderson localization due to dephasing.
This implies that the conducting channels are balanced
as a consequence of the mixing of two valleys
caused by strong intervalley scattering, in contrast to the LRI case.
In the absence of dephasing, $\langle g\rangle$ decays
exponentially with increasing $L/a$ as shown in Fig.~\ref{fig:SRI_semilog}.
We show error bars only in Fig.~\ref{fig:SRI_semilog}
as they are very small in the case with dephasing.

\section{Comparison with the Boltzmann Transport Theory}

In this section, we introduce an analytical expression for the dimensionless
conductance derived from the Boltzmann transport equation~\cite{takane2}
and compare it with the numerical results given in the previous section.

The analytical expression of Ref.~\citen{takane2} is derived by applying
the Boltzmann transport equation to a simple model for zigzag nanoribbons
and hence is justified
in the incoherent limit where the phase coherence of electrons is completely lost.
The model possesses two energy valleys, and the number of conducting channels
for right-going (left-going) electrons is $N+1$ ($N$) in one valley,
while in the other valley, that for right-going (left-going) electrons
is $N$ ($N+1$).
The total number of conducting channels is given by $N_{\text c} = 2N+1$
including contributions from the two valleys.
Disorder induces intravalley scattering between two channels in the same valley
and intervalley scattering between two channels belonging to
different valleys.
The strength of intravalley scattering is characterized by a single parameter
$\kappa$ as its detailed dependence on initial and final states is ignored.
In the same manner, the strength of intervalley scattering is characterized by a parameter
$\kappa'$.
Applying a constant electric field only in the region of length $L$,
the Boltzmann equation is solved under the condition that incident electrons
from the left and right are described by equilibrium distributions.
The resulting dimensionless conductance $g_{\rm B}$ is expressed as~\cite{takane2}
\begin{align}
  g_{\rm B}
    = \frac{\left[\kappa + (8N^{2}+8N+1)\kappa' \right]c_{L}}
           {(2N+1)\kappa'\left[2+(2N+1)(\kappa+\kappa')L
                          \right]c_{L}
            + \frac{\kappa^{2}-{\kappa'}^{2}}{\sqrt{\alpha}}d_{L}} ,
	\label{gb}
\end{align}
where $\alpha=(\kappa+\kappa')[\kappa+(8N^{2}+8N+1)\kappa']$ and
\begin{align}
  c_{L} & = \frac{(2N+1)(\kappa+\kappa')}{\sqrt{\alpha}}
            \cosh(\sqrt{\alpha}L/2) + \sinh(\sqrt{\alpha}L/2) ,
               \\
  d_{L} & = \frac{(2N+1)(\kappa+\kappa')}{\sqrt{\alpha}}
            \sinh(\sqrt{\alpha}L/2) + \cosh(\sqrt{\alpha}L/2) .
\end{align}

\begin{figure}[htpb]
	\centering
	\includegraphics{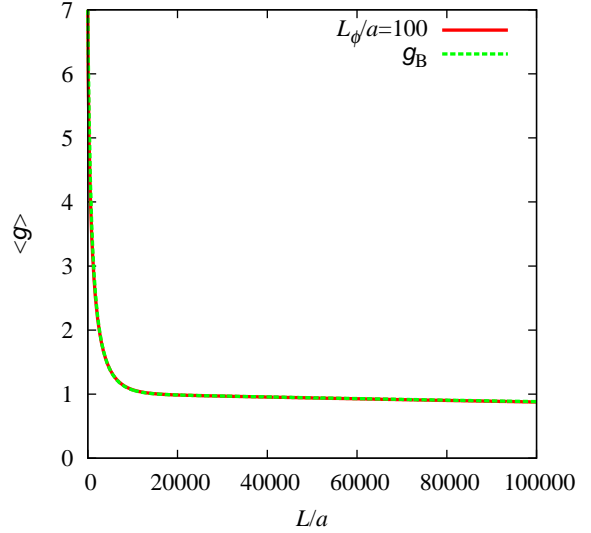}
	\caption{
	(Color online)
	Fitting of $\langle g\rangle$ at $L_{\phi}/a=100$
	in the LRI case with the analytical expression $g_{\text B}$.
	}
	\label{fig:fitting_LRI}
\end{figure}

\begin{figure}[htpb]
	\centering
	\includegraphics{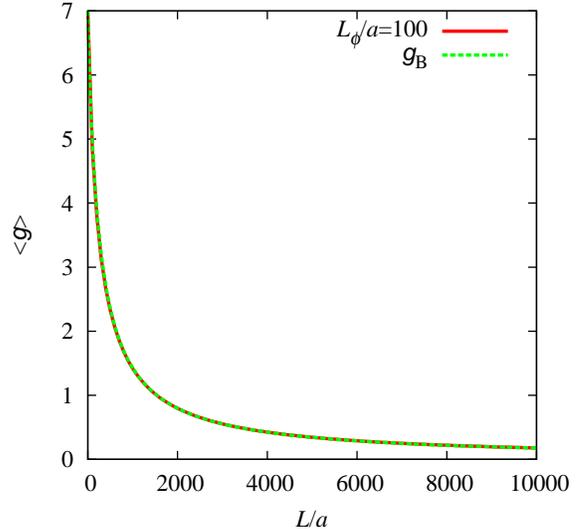}
	\caption{
	(Color online)
	Fitting of $\langle g\rangle$ at $L_{\phi}/a=100$
	in the SRI case with the analytical expression $g_{\text B}$.
	}
	\label{fig:fitting_SRI}
\end{figure}

We examine whether this expression can fit our numerical results.
Equation (\ref{gb}) is justified in the incoherent limit, so we focus on the numerical results
in the smallest-$L_{\phi}$ case with $L_{\phi}/a = 100$.
As $N_{\text c}=7$ in our setting, $N$ is fixed at $N = 3$.
Only $\kappa$ and $\kappa'$ play the role of fitting parameters.
The result of fitting is shown in Fig.~\ref{fig:fitting_LRI} in the LRI case
and Fig.~\ref{fig:fitting_SRI} in the SRI case.
We observe that the analytical expression accurately reproduces
the numerical results.
The best fitting is achieved for $\kappa a = 0.00031$ and $\kappa' a = 0.00000003$ in the former case,
yielding $\kappa'/\kappa = 0.0000096$,
and for $\kappa a = 0.00094$ and $\kappa' a = 0.00021$ in the latter case,
yielding $\kappa'/\kappa = 0.223$.
This result clearly indicates that intervalley scattering is significantly
weaker than intravalley scattering in the LRI case
while their strengths are on the same order of magnitude in the SRI case.

One may think that the effect of dephasing is
oversimplified in the model used in our numerical calculations.
However, the above result implies that our model captures the essential features
of dephasing in spite of its simplicity.


\section{Summary and Conclusion}
We have studied the effect of dephasing
on a perfectly conducting channel (PCC) 
in disordered graphene nanoribbons with zigzag edges
by numerically calculating
the average dimensionless conductance $\langle g\rangle$ as a fucntion
of system length $L$.
We separately consider the case with long-range impurities (LRIs) and
that with short-range impurities (SRIs).
In the former case, intervalley scattering is
very weak and the appearance of a PCC is expected.
Contrastingly, a PCC cannot appear owing to
strong intervalley scattering in the latter case.
The result of the LRI case
indicates that $\langle g\rangle$ as a function of $L$ shows two-stage
behavior;
$\langle g\rangle$ rapidly decreases to 1 with increasing $L$ in the first stage
and then tends to decay below 1 in the second stage.
The behavior in the first stage implies the presence of a PCC,
and the behavior in the second stage indicates that the PCC is destabilized
by weak intervalley scattering.
We have clearly observed that
dephasing significantly relaxes the second-stage behavior 
and hence effectively stabilizes the PCC.
This stabilization should be attributed to the suppression of
Anderson localization due to dephasing.
In the SRI case,
$\langle g\rangle$ decays toward zero, reflecting the absence of a PCC.
We have shown that
dephasing suppresses the effect of Anderson localization,
revealing the behavior of $\langle g\rangle \propto L^{-1}$
in accordance with the ordinary Ohm's law.

One may think that the experimental detection of a PCC in realistic systems
is not easy as various inelastic processes obstruct it.
Among them, dephasing is known as the most notable factor that suppresses
quantum behaviors
of electrons at low temperatures.
Indeed, it has been pointed out that
a PCC in CNTs, as well as in topological insulators, 
is fragile against dephasing.\cite{ando3}
Contrastingly,
in graphene nanoribbons,
dephasing does not negatively influence a PCC
but rather encourages its appearance.
This implies that
graphene nanoribbons with zigzag edges
are a promising platform for the experimental detection of a PCC.

\section*{Acknowledgment}

This work was supported by a Grant-in-Aid for Scientific Research (C)
(No. 15K05130).

\end{document}

%% file: 67329Fig3.pstex_t
\begin{picture}(0,0)%
\includegraphics{67329Fig3.pstex}%
\end{picture}%
\setlength{\unitlength}{4144sp}%
\begingroup\makeatletter\ifx\SetFigFont\undefined%
\gdef\SetFigFont#1#2#3#4#5{%
  \reset@font\fontsize{#1}{#2pt}%
  \fontfamily{#3}\fontseries{#4}\fontshape{#5}%
  \selectfont}%
\fi\endgroup%
\begin{picture}(3713,2515)(3,-1880)
\put(1621,-376){\makebox(0,0)[lb]{\smash{{\SetFigFont{10}{12.0}{\rmdefault}{\mddefault}{\updefault}{\color[rgb]{0,0,0}${\cal R}_1$}%
}}}}
\put(2341,-376){\makebox(0,0)[lb]{\smash{{\SetFigFont{10}{12.0}{\rmdefault}{\mddefault}{\updefault}{\color[rgb]{0,0,0}${\cal T}_2$}%
}}}}
\put(2341,-556){\makebox(0,0)[lb]{\smash{{\SetFigFont{10}{12.0}{\rmdefault}{\mddefault}{\updefault}{\color[rgb]{0,0,0}${\cal T}_{\overline{2}}$}%
}}}}
\put(2656,-556){\makebox(0,0)[lb]{\smash{{\SetFigFont{10}{12.0}{\rmdefault}{\mddefault}{\updefault}{\color[rgb]{0,0,0}${\cal R}_{\overline{2}}$}%
}}}}
\put(2656,-376){\makebox(0,0)[lb]{\smash{{\SetFigFont{10}{12.0}{\rmdefault}{\mddefault}{\updefault}{\color[rgb]{0,0,0}${\cal R}_2$}%
}}}}
\put(1801,-1636){\makebox(0,0)[lb]{\smash{{\SetFigFont{10}{12.0}{\rmdefault}{\mddefault}{\updefault}{\color[rgb]{0,0,0}${\cal T}_{N_s}$}%
}}}}
\put(1801,-1816){\makebox(0,0)[lb]{\smash{{\SetFigFont{10}{12.0}{\rmdefault}{\mddefault}{\updefault}{\color[rgb]{0,0,0}${\cal T}_{\overline{N}_s}$}%
}}}}
\put(2116,-1816){\makebox(0,0)[lb]{\smash{{\SetFigFont{10}{12.0}{\rmdefault}{\mddefault}{\updefault}{\color[rgb]{0,0,0}${\cal R}_{\overline{N}_s}$}%
}}}}
\put(2116,-1636){\makebox(0,0)[lb]{\smash{{\SetFigFont{10}{12.0}{\rmdefault}{\mddefault}{\updefault}{\color[rgb]{0,0,0}${\cal R}_{N_s}$}%
}}}}
\put(1306,-376){\makebox(0,0)[lb]{\smash{{\SetFigFont{10}{12.0}{\rmdefault}{\mddefault}{\updefault}{\color[rgb]{0,0,0}${\cal T}_1$}%
}}}}
\put(1306,-556){\makebox(0,0)[lb]{\smash{{\SetFigFont{10}{12.0}{\rmdefault}{\mddefault}{\updefault}{\color[rgb]{0,0,0}${\cal T}_{\overline{1}}$}%
}}}}
\put(1531,524){\makebox(0,0)[lb]{\smash{{\SetFigFont{8}{9.6}{\rmdefault}{\mddefault}{\updefault}{\color[rgb]{0,0,0}$L_\phi$}%
}}}}
\put(2566,524){\makebox(0,0)[lb]{\smash{{\SetFigFont{8}{9.6}{\rmdefault}{\mddefault}{\updefault}{\color[rgb]{0,0,0}$L_\phi$}%
}}}}
\put(1981,-736){\makebox(0,0)[lb]{\smash{{\SetFigFont{8}{9.6}{\rmdefault}{\mddefault}{\updefault}{\color[rgb]{0,0,0}$L_\phi$}%
}}}}
\put(1621,-556){\makebox(0,0)[lb]{\smash{{\SetFigFont{10}{12.0}{\rmdefault}{\mddefault}{\updefault}{\color[rgb]{0,0,0}${\cal R}_{\overline{1}}$}%
}}}}
\end{picture}%